\documentclass[]{spie}  

\usepackage{amsmath,amsfonts,amssymb}
\usepackage{graphicx}
\usepackage{xcolor}
\usepackage[colorlinks=true, allcolors=blue]{hyperref}
\usepackage{cleveref}
\usepackage{subcaption}
\usepackage{float}
\usepackage{tikz}
\usetikzlibrary{shapes.geometric,arrows.meta,fit,calc,positioning}

\crefname{section}{sec.}{secs.}
\Crefname{section}{Sec.}{Secs.}

\crefname{figure}{fig.}{figs.}
\Crefname{figure}{Fig.}{Figs.}

\title{CCAT: a two-octave 1.024\,GHz KID readout featuring an
overlap-channel polyphase synthesis filter bank on RFSoC}

\author[a,*]{Ruixuan (Matt) Xie}
\author[b]{Adrian K. Sinclair}
\author[a]{James Burgoyne}
\author[d]{Maxwell Chapman}
\author[c,a]{Scott Chapman}
\author[e]{Steve K. Choi}
\author[a]{Anthony I. Huber}
\author[f]{Ben Keller}
\author[f,g]{Michael D. Niemack}

\affil[a]{Dept.\ of Physics and Astronomy,
          University of British Columbia, Vancouver, BC, Canada}
\affil[b]{The William H.\ Miller III Department of Physics and Astronomy,
          Johns Hopkins University, Baltimore, MD, United States}
\affil[c]{Dept.\ of Physics and Atmospheric Science,
          Dalhousie University, Halifax, NS, Canada}
\affil[d]{Dept.\ of Mechanical Engineering, McGill University, Montreal, QC, Canada}
\affil[e]{Center for Experimental Cosmology and Instrumentation, Dept.\ of Physics and Astronomy, University of California, Riverside, CA 92521, USA}
\affil[f]{Dept.\ of Physics, Cornell University, Ithaca, NY, 14853, USA}
\affil[g]{Dept.\ of Astronomy, Cornell University, Ithaca, NY, 14853, USA}

\begin{document}
\maketitle

\begin{abstract}
Next-generation submillimeter instruments require gigahertz-scale readout bandwidths to support the growing detector counts of large microwave kinetic inductance detector (KID) arrays. Designed to meet the readout bandwidth and tone-capacity requirements of the CCAT Prime-Cam 850\,GHz and 410\,GHz instrument modules, we developed our second-generation (Gen2) KID readout system on a Xilinx ZCU111 radio frequency system-on-chip (RFSoC), based on a parallelized overlap-channel polyphase synthesis filter bank and a companion wideband receiving channelizer. This two-octave architecture reads out four independent RF networks, each with 1.024\,GHz instantaneous bandwidth and up to 2048 detectors. The polyphase synthesis doubles the baseline channel count and bandwidth, and enables on-the-fly control of individual tone frequency, amplitude and phase, facilitating optimized KID biasing and providing a gateway to tone tracking. These capabilities may also be useful more broadly across Prime-Cam and in similar frequency-division multiplexed (FDM) readout systems. We present the design in DSP simulation, and measurements in RF loopback as well as with two KID test chips, one spanning the full two-octave band. Loopback measurements demonstrate comparable noise performance between the first-generation (Gen1) baseline and Gen2; preliminary on/off-resonance measurements indicate detector-noise-limited operation for the majority of channels; and the digital channel crosstalk is measured and discussed. We report FPGA resource and power utilization and assess scalability toward future large-format KID instruments.
\end{abstract}


{\noindent \footnotesize\textbf{*}Correspondence should be addressed to R.X.:
\href{mailto:MattXie956@gmail.com}{MattXie956@gmail.com}}

\section{INTRODUCTION}
\label{sec:intro}

Observations at submillimeter and far-infrared wavelengths probe some of the most anticipated and informative processes in the Universe, including galaxy assembly through the epoch of reionization, large-scale structure through the Sunyaev--Zeldovich effect, cosmic star formation, and dust-polarized foregrounds for CMB $B$-mode measurements.\cite{CCAT_Prime_Collaboration_2022} The scientific return from these observations scales strongly with mapping speed, which in turn scales with detector count. Microwave kinetic inductance detectors (KIDs) have emerged as a technology of choice for large-format sub-mm focal planes, because of their high multiplexing factor, photon-noise-limited sensitivity~\cite{vaskuri2025}, fabrication simplicity, and natural compatibility with frequency-division multiplexing (FDM) readout~\cite{sinclair2022}. In FDM, thousands of KIDs, each a superconducting LC resonator with a unique resonant frequency, are capacitively coupled to a single coaxial transmission line and simultaneously read out via a wideband frequency comb of probe tones (also called bias tones).\cite{xie2024,xie2024thesis}

As detector counts grow, the readout bottleneck has shifted from cryogenic to warm electronics: the limiting factor is no longer cryogenic wiring but warm-side radio-frequency data converter (RFDC) and digital signal processing (DSP) that must synthesize and analyze ever-wider frequency combs.\cite{xie2024,xie2024thesis} Because each KID resonance has a finite linewidth and adjacent resonators require sufficient frequency spacing to limit collisions and crosstalk, increasing the detector count generally requires a proportionally larger readout bandwidth. Radio-frequency systems-on-chips (RFSoCs) address this by integrating high-speed data converters (ADCs and DACs) with programmable FPGA fabric and an integrated processing system on a single die, providing the bandwidth density and power efficiency that modern KID readouts demand. The RFSoC and FPGA technology has enabled a new generation of compact, gigahertz-bandwidth KID readout systems and is now a standard development platform~\cite{sinclair2022,xie2024thesis,smith2024_mkidgen3,lit_rev_OSPFBonRedPitaya_Arnaldi_2021,bryan2026HWO_readout,basha2026HWO_pfb,Essinger_Hileman2026PRIMA_readout}.

The Fred Young Submillimeter Telescope (FYST),\cite{CCAT_Prime_Collaboration_2022} located at 5600\,m on Cerro Chajnantor in the Chilean Atacama Desert, will host the Prime-Cam instrument, whose frequency modules span 280\,GHz to 850\,GHz and together field over 100,000 KIDs when its seven instrument modules are fully populated. The 850\,GHz broadband instrument module~\cite{Chapman850GHz2022} presents the most demanding readout requirements: approximately 38,000 polarization-sensitive KIDs distributed across three arrays with 12 RF networks each, each network designed with 1052 or 1056 resonators, of which up to 1024 are targeted for readout~\cite{burgoyne2026ledmapping}. A novel inductance shorting technique is developed to manage multi-octave KID distributions in the 850\,GHz science-grade arrays, which are currently under fabrication and characterization~\cite{huberA2024,huberA2026}. In one RF network, the resonances span approximately 0.2–1.2\,GHz and are covered in a single acquisition by the 1.024\,GHz instantaneous bandwidth provided by the readout system developed in this work. The 410\,GHz instrument module (recently funded and in development)~\cite{patel2026opticaldesign,chapman2026} shares the same readout bandwidth requirement.

A baseline RFSoC readout, hereafter the first-generation (Gen1) readout, developed by Sinclair et al.~\cite{sinclair2022,sinclair2023,sinclair2024noise}, is currently deployed and operational with the Prime-Cam and Mod-Cam laboratory test instrument~\cite{patel2025ccatreadout10000280}. In this system, the frequency comb is provided by a waveform look-up table (LUT) stored in DDR4 memory and streamed to the DAC. Although effective, the LUT approach constrains the instantaneous bandwidth to 512\,MHz per RF network because of DDR4 memory depth and data-transfer-rate limitations, and does not support real-time adjustment of the waveform. The LUT approach is sufficient for the lower-frequency Prime-Cam modules in terms of bandwidth requirements. For the 850\,GHz and 410\,GHz modules, however, the 512\,MHz bandwidth is insufficient due to their larger detector counts per network. The runtime waveform editing capability is advantageous for all modules.

In 2024,\cite{xie2024,xie2024thesis} we demonstrated an overlap-channel polyphase synthesis filter bank (OC-PSB) implemented on a Xilinx ZCU111 RFSoC, synthesizing probe tones in real time and overcoming the constraints of the baseline LUT approach. The prototype ran without datapath parallelism, at 256\,MHz FPGA clock rate yielding a 256\,MHz instantaneous bandwidth, and showcased the OC-PSB in isolation on the transmit path only. Its 2048-channel structure was, however, designed from the outset to reach the full 1.024\,GHz band by processing four samples per clock.

The present paper extends the 2024 OC-PSB to a complete second-generation (Gen2) KID readout system: four independent readout chains, each comprising the OC-PSB transmitter paired with a wideband FFT channelizer receiver. Relative to Gen1, Gen2 doubles the instantaneous bandwidth per RF network (1.024\,GHz versus\ 512\,MHz) and supports up to 8192 probe tones per RFSoC across the four chains. The OC-PSB enables per-tone runtime control of frequency, amplitude and phase. This capability is a prerequisite for ``tone tracking,'' in which probe tones follow shifting KID resonance frequencies during observations. Gen2 is planned for deployment with the 850\,GHz and 410\,GHz Prime-Cam instrument modules and has been tested on two test chips, one of which spans two octaves of bandwidth, enabling a full-band sweep that Gen1 cannot perform. This paper presents the Gen2 DSP architecture, implementation, and measured performance.

The remainder of this paper is organized as follows. \Cref{sec:overview} provides a system overview. \Cref{sec:dsp} details the Gen2 DSP signal chain upgrades. \Cref{sec:implementation} covers FPGA implementation and resource utilization. \Cref{sec:results} presents measured performance. \Cref{sec:conclusion} summarizes and outlines future work.

\clearpage
\section{SYSTEM OVERVIEW}
\label{sec:overview}

\begin{figure}[h]
    \centering
    \includegraphics[width=\textwidth]{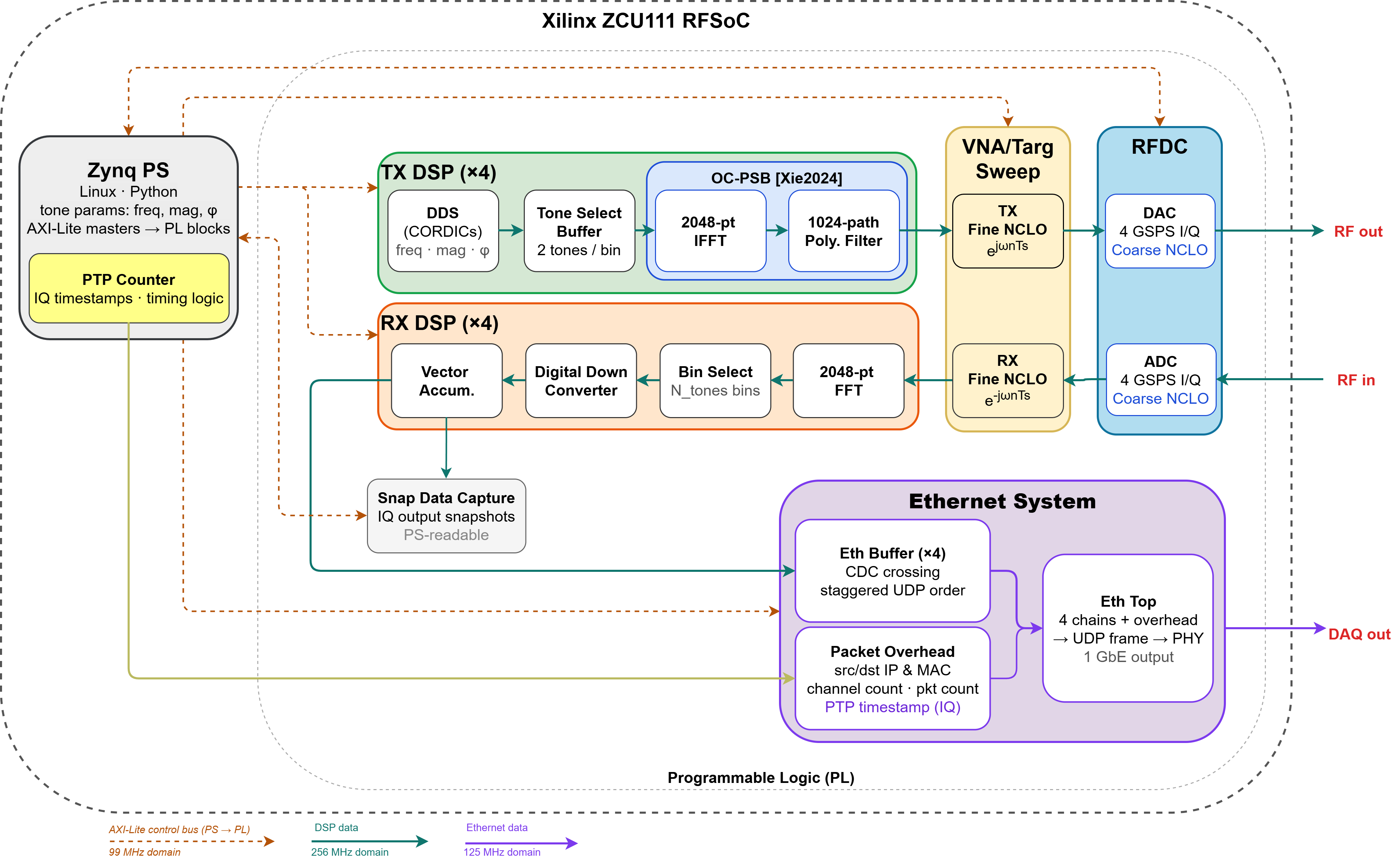}
    \vspace{2pt}
    \caption{Block diagram of the Gen2 readout system on the Xilinx ZCU111 RFSoC. Four independent readout chains are implemented, with one shown in detail. \emph{Transmit path (TX\,DSP):} a direct digital synthesizer (DDS) driven by per-tone frequency, magnitude, and phase parameters from the Zynq processing system (PS) feeds a Tone-Select buffer and the overlap-channel polyphase synthesis filter bank (OC-PSB\,\cite{xie2024,xie2024thesis}), which comprises a 2048-point IFFT followed by a 1024-path polyphase filter. A fine numerically controlled local oscillator (NCLO) after the OC-PSB shifts the entire comb in frequency for VNA sweeps before reaching the DAC. \emph{Receive path (RX\,DSP):} after ADC sampling, a conjugate fine NCLO preceding the 2048-point FFT cancels the TX sweep so that the downstream bin-select, digital downconverter, and vector accumulator always operate on a stationary digital comb. A snapshot BRAM buffer branches from the vector accumulator to provide PS-accessible captures of complex output samples (in-phase and quadrature, or I/Q). The Ethernet system collects output data from all four chains via per-chain clock-domain-crossing (CDC) buffers, adds source and destination IP and MAC addresses together with the channel count, packet count, and PTP timestamp, and emits UDP packets over 1\,GbE to the downstream data acquisition. Programmable-logic blocks are configured via AXI-Lite from the PS. Precision Time Protocol (PTP) is used to timestamp the output I/Q samples.}
    \label{fig:gen2_arch}
\end{figure}

The overall design of the Gen2 system is derived from the Gen1 system developed by Sinclair et al.~\cite{sinclair2022}, with substantial upgrades to the DSP of the transmitter and a reimplemented simplified receiver. Supporting modules, including the fine NCLOs, Ethernet subsystem, and PTP timestamping, are largely adapted from their Gen1 counterparts, with modifications to accommodate Gen2 operating conditions and system requirements.

\Cref{fig:gen2_arch} shows the Gen2 system architecture. The Gen2 readout is implemented on a Xilinx ZCU111 evaluation board hosting a Zynq UltraScale+ RFSoC device. Four independent readout chains are implemented, each served by a dedicated DAC--ADC pair in the RF data converter (RFDC). Each chain provides 1.024\,GHz of instantaneous bandwidth centered at a numerically controlled local oscillator (NCLO) frequency that can be set anywhere within the RFDC's usable range. An Opsero Ethernet FMC provides Gigabit Ethernet ports for detector data streaming and board control. The overall readout operation, including VNA-sweep resonator identification, tone assignment, and data streaming, follows the Gen1 operational procedures detailed by Patel et al.~\cite{patel2025ccatreadout10000280}. Gen2 aims to be software-compatible with the corresponding Gen1 infrastructure.

The FPGA fabric operates at 256\,MHz clock rate with a parallel-by-four architecture, yielding an effective processing rate of 1.024\,GHz per chain. Each chain produces complex output samples (in-phase and quadrature, or I/Q) at 488.28\,Hz (variable in software), packetized as UDP datagrams and timestamped with IEEE\,1588 Precision Time Protocol (PTP), giving ${\approx}16$\,MB/s of detector payload per board across four chains.

\Cref{tab:gen2_vs_gen1} summarizes the key capability differences between Gen1 and Gen2.

\begin{table}[H]
\centering
\caption{Gen1 versus\ Gen2 readout capability comparison.}
\label{tab:gen2_vs_gen1}
\vspace{6pt}
\begin{tabular}{lll}
\hline
Parameter & Gen1\cite{sinclair2022,sinclair2024noise} & Gen2 (this work) \\
\hline
Instantaneous BW per chain & 512\,MHz & 1.024\,GHz \\
Tone synthesis method       & DDR4 waveform LUT & OC-PSB (FPGA fabric) \\
Per-tone runtime control    & No  & Yes (freq., amp., phase) \\
Max tones per chain         & $\sim$1000 & 2048 \\
Max tones per board         & $\sim$4000 & 8192 \\
Receive channelizer         & PFB + FFT & FFT only \\
RF chains per board         & 4 & 4 \\
\hline
\end{tabular}
\end{table}

\section{DSP SIGNAL CHAIN}
\label{sec:dsp}

\subsection{OC-PSB transmitter}
\label{sec:dsp:tx}

The Gen2 transmitter replaces the Gen1 DDR4 waveform LUT with the overlap-channel polyphase synthesis filter bank (OC-PSB), as presented and detailed previously~\cite{xie2024,xie2024thesis}. The OC-PSB is a parallelized FPGA-fabric architecture that synthesizes up to 2048 probe tones across the full 1.024\,GHz band in real time. Tones are generated by DDS (CORDIC operating in rotation mode), from runtime-writable per-tone state: an initial vector sets the tone's amplitude and initial phase, and a per-clock-cycle phase increment sets its frequency.

The OC-PSB produces 2048 channels of 1\,MHz bandwidth with 50\,\% overlap (500\,kHz channel spacing). The effective tone placement resolution is about 15\,Hz, which is the synthesizer's native placement resolution (limited by the fixed-point precision of phase increments). In the complete readout, the tones are additionally snapped to the 488.28125\,Hz null grid detailed in \Cref{sec:dsp:rx}, so the effective system-level frequency resolution is 488.28125\,Hz. The 50\,\% overlap between adjacent synthesis channels is the key property that eliminates inter-channel amplitude roll-off, allowing tones to be placed anywhere in the band with uniform response.

Beyond bandwidth and flexibility, the runtime per-tone control enabled by the OC-PSB is a prerequisite for tone tracking, in which the probe tone frequency tracks that of the KID resonance during an observation; the frequency of the tracking tone carries the photon-power information directly, skipping the translation from phase to frequency shift.

\Cref{tab:ocpsb_params} summarizes the resulting parallelized OC-PSB performance. Some entries differ from the pre-parallelism results presented in 2024~\cite{xie2024,xie2024thesis}. These results characterize the transmitter in isolation (with the synthesized comb measured at the DAC output with a spectrum analyzer), distinct from the readout noise reported in \Cref{sec:results:noise}.

\begin{table}[h]
\centering
\caption{Measured Gen2 OC-PSB transmitter performance. $^{\dagger}$One tone within a 1024-tone comb, measured on a spectrum analyzer whose own noise floor ($-110$\,dBm at 1\,kHz RBW) sets the measurement limit; the transmitter's true noise floor lies below this and is unresolved.}
\label{tab:ocpsb_params}
\vspace{6pt}
\begin{tabular}{ll}
\hline
Parameter & Value \\
\hline
Synthesized bandwidth        & 1.024\,GHz \\
Synthesis channels (IFFT length) & 2048 \\
Channel bandwidth            & 1\,MHz \\
Channel overlap              & 50\,\% \\
Tone placement resolution    & ${{\sim}}15$\,Hz \\
Max tones per chain          & 2048 \\
Parallelism factor           & 4 (versus\ 1 in Refs.~\citenum{xie2024,xie2024thesis}) \\
Fabric clock                 & 256\,MHz \\
SNR (floating-point sim)     & $\geq 110$\,dB \\
SNR, 1\,MHz offset (measured) & $\geq 83.39$\,dB$^{\dagger}$ \\
\hline
\end{tabular}
\end{table}

\subsection{Tone-Select reallocation}
\label{sec:dsp:toneselect}
\begin{figure}[htbp]
  \centering
  \includegraphics[width=0.8\textwidth]{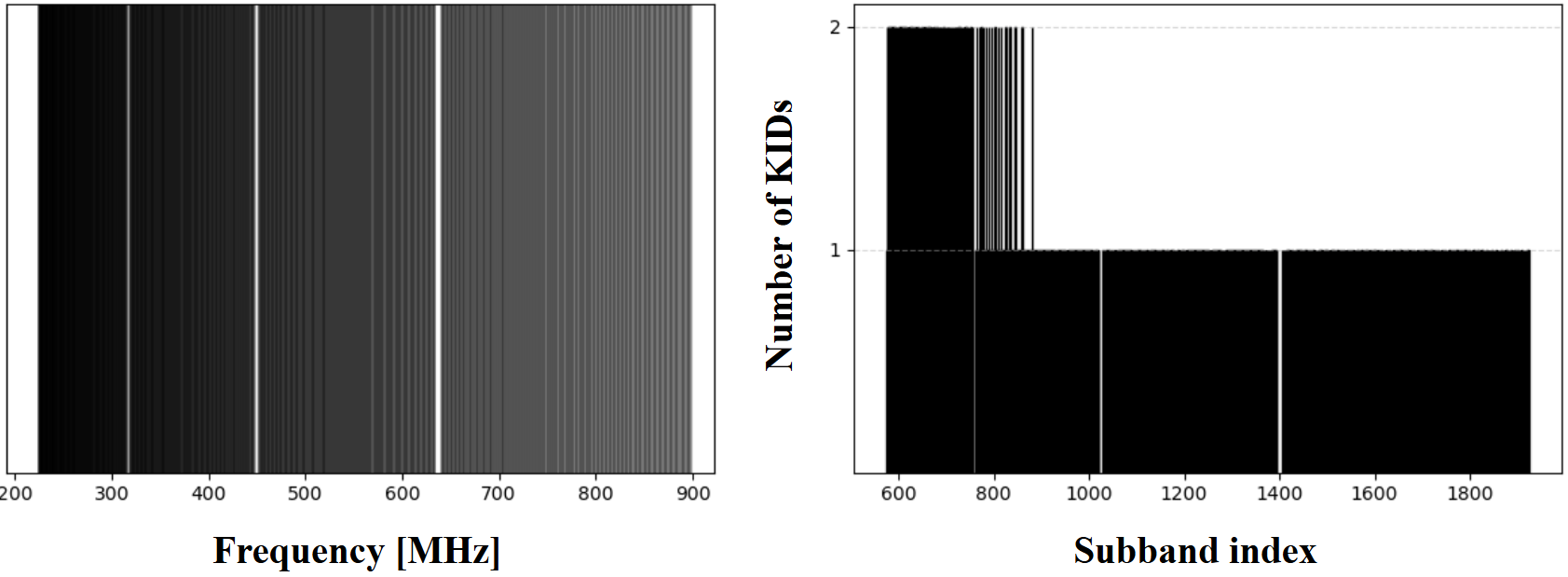}
  \vspace{2pt}
  \caption{Preliminary scheduled frequency distribution and synthesis channel occupancy for one RF network of the 850\,GHz KID array. This preliminary schedule contains 1060 resonators and predates the current 1052/1056-resonator network design. Of the 2048 available channels, 1086 are empty, 864 contain one KID, and 98 contain two KIDs, yielding 962 occupied channels and 1060 scheduled resonance frequencies in total. No channel contains more than two resonators.
  \emph{Left:} Each vertical line represents one scheduled KID resonance; the vertical position has no physical meaning. The greater line density at lower frequencies shows that the resonators are more closely packed in this region.
  \emph{Right:} Channel occupancy as a function of subband index. Doubly occupied channels occur predominantly at the lower-frequency region of the array, while the higher-frequency region contains only singly occupied channels.}
  \label{fig:scheduled_frequency_occupancy}
\end{figure}

The OC-PSB combines time-division multiplexed (TDM) input subband channels and outputs an FDM full-band waveform. And the TDM CORDICs in rotation mode natually supply one tone per synthesis channel (IFFT bin)~\cite{xie2024thesis}. The detectors are designed to be distributed across the available bandwidth, so that most occupy distinct synthesis channels. Fabrication variations can nevertheless place multiple detectors within the same 500\,kHz channel spacing. Generally in densely packed KID arrays, such instances grow more common toward the low-frequency end of the band and arise from the array design and fabrication process.
\Cref{fig:scheduled_frequency_occupancy} shows the frequency distribution and synthesis channel occupancy of the preliminary KID frequencies scheduled for an RF network in our 850\,GHz KID array. A non-negligible number of channels contain two resonators and thus a two tone per channel solution is required to read out all the resonators.

With up to 1024 resonators targeted for readout per network, but 2048 available synthesis channels, about half the tone slots would otherwise go unused. While enabling doubly occupied channels with a duplicate DDS path roughly doubles the number of CORDICs employed, Gen2 instead introduces a Tone-Select stage between the DDS and the 2048-point IFFT, reallocating the capacity that would otherwise drive an empty channel to supply the second tone of a doubly occupied channel. Every channel is thus capable of accommodating a second tone while none is dedicated a priori to holding two. For the analyzed schedule in \Cref{fig:scheduled_frequency_occupancy}, the 1086 vacant channels substantially outnumber the 98 doubly occupied channels, leaving sufficient spare tone slots for all required reallocations. The absence of triply occupied channels in the analyzed schedule makes the two-tone ceiling sufficient for the intended array. Channels containing three or more resonators are treated as collisions, where resonator responses overlap in frequency space. In this case, two well-separated resonators are retained while the rest are excluded from readout.

The Tone-Select mapping is determined at probe tone comb setup time in software. A VNA frequency sweep~\cite{patel2025ccatreadout10000280} locates the resonators and classifies each channel as holding zero, one, or two KIDs, populating a LUT that drives Tone-Select. The TDM tone signals from DDS originally arrive in nominal channel order that a streaming IFFT expects; Tone-Select buffers them and, for each doubly occupied channel, adds the tone generated for an otherwise-empty channel to that channel's primary tone before the sum arrives at the IFFT, while single-tone channels pass through unmodified. This is inspired by, but distinct from, the receiver Bin-Select stage which recovers the two tones of a shared channel independently. The summation is implemented with a single two-input adder per path, so at most one extra tone can be accommodated per channel by the Tone-Select. The 50\% overlap in the OC-PSB provides another, yet-unexplored pathway for placing tones within a channel.

\subsection{Receive channelizer}
\label{sec:dsp:rx}
The Gen2 receive path replaces the Gen1 polyphase analysis filter bank (PFB) with a 2048-point FFT channelizer, thereby removing the polyphase-filter front end. The PFB serves two functions: inter-bin alias suppression and passband flattening. We argue that, under appropriate constraints, both functions can be achieved without a PFB.

\textbf{Alias suppression.}
Without a PFB, each FFT bin has the frequency response of a rectangular window, which provides poor stopband rejection. A probe tone that is not located exactly at a bin center therefore produces nonzero responses in multiple FFT bins, and these out-of-bin components can alias into the retained bin bandwidth after decimation. The exceptional case is a tone exactly aligned with the 500\,kHz FFT-bin grid, for which the sampled FFT response is confined to a single bin. In practice, KID resonance frequencies are set by fabrication and occur at arbitrary locations across the readout band, so the probe tones are generally off-grid. Even with ideal fabrication, the resonances shift under varying sky loading conditions. A plain FFT channelizer therefore cannot provide adequate channel isolation for arbitrary and time-varying tone locations. 

Following the FFT is a Bin-Select stage, which selects only the output bins that contain probe tones, skipping the empty ones and supplying a doubly occupied bin twice (so the nominal bin order $0,1,2,3,\ldots$ becomes a selected sequence such as $2,4,7,7,8,\ldots$). Alias suppression is recovered downstream during averaging.

Each selected bin is digitally down-converted (DDC) to baseband, then averaged over \(N_{\mathrm{avg}}=1024\) consecutive samples and decimated by the same factor. A shared bin's two reads use different DDC frequencies, so its two co-resident tones are recovered independently. Operating at the per-bin rate $f_\text{bin}=500$\,kHz, the averaging filter has the Dirichlet magnitude response
\begin{equation}
  \left| H(f) \right|
    = \frac{1}{N_\text{avg}}
      \left| \frac{\sin\!\left(\pi N_\text{avg}\, f / f_\text{bin}\right)}
                   {\sin\!\left(\pi f / f_\text{bin}\right)} \right|,
  \label{eq:boxcar}
\end{equation}
where \(f\) is the frequency offset from the DDC-centered carrier. The transmission nulls occur wherever $N_\text{avg}\, f / f_\text{bin}$ is a non-zero integer, i.e., at integer multiples of
\begin{equation}
  \Delta f_\text{null}
    = \frac{f_\text{bin}}{N_\text{avg}}
    = \frac{500\,\text{kHz}}{1024}
    = \frac{1.024\,\text{GHz}}{2048 \times 1024}
    = 488.28125\,\text{Hz}.
  \label{eq:null_spacing}
\end{equation}
The null spacing is set by the sampling rate $f_\text{bin}$ and the accumulation length $N_\text{avg}$, and it is also the per-tone output sample rate, so each carrier's surviving output band is only $\pm f_\text{bin}/(2 N_\text{avg}) \approx \pm244$\,Hz wide. By snapping every probe tone, in software, to the nearest multiple of $\Delta f_\text{null}$ before synthesis, we ensure that the aliases from spectral folding lands on a boxcar null where it is rejected.

\begin{figure}[p]
    \centering

    \begin{subfigure}{\textwidth}
        \centering
        \includegraphics[width=0.9\linewidth]{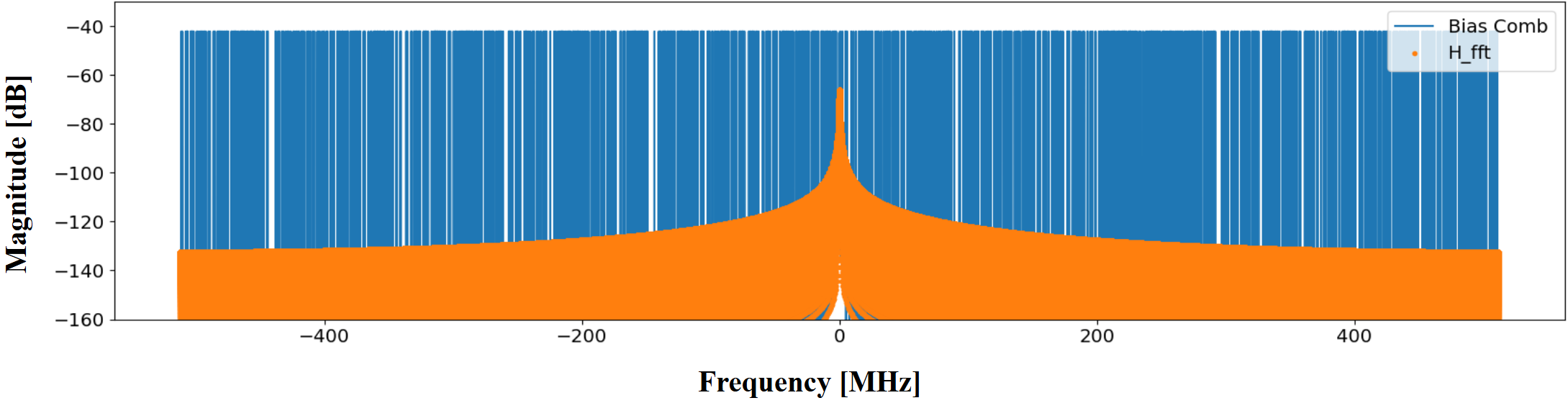}
        \caption{Input comb across the full 1.024\,GHz band (centered at DC in the digital domain) with the single-FFT-bin frequency response overlaid.}
        \label{fig:no_pfb_sim-a}
    \end{subfigure}

    \begin{subfigure}{\textwidth}
        \centering
        \includegraphics[width=0.9\linewidth]{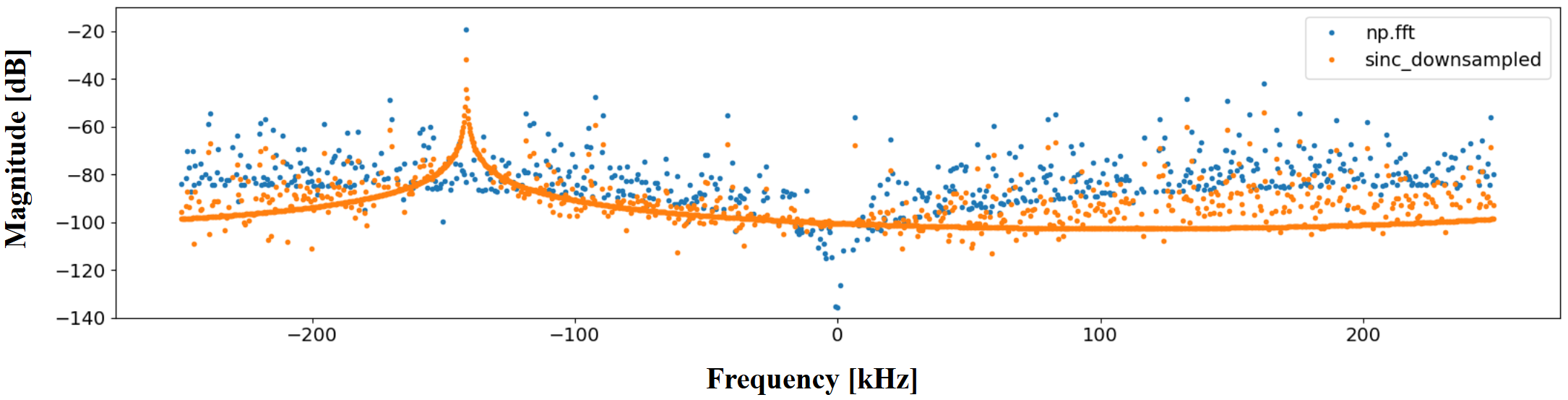}
        \caption{FFT-bin output spectrum (500\,kHz sub-band), computed both directly by FFT of the time series and by multiplying the input comb with the analytic $\operatorname{sinc}$ bin response. The two agree, validating the frequency-domain model.}
        \label{fig:no_pfb_sim-b}
    \end{subfigure}

    \begin{subfigure}{\textwidth}
        \centering
        \includegraphics[width=0.9\linewidth]{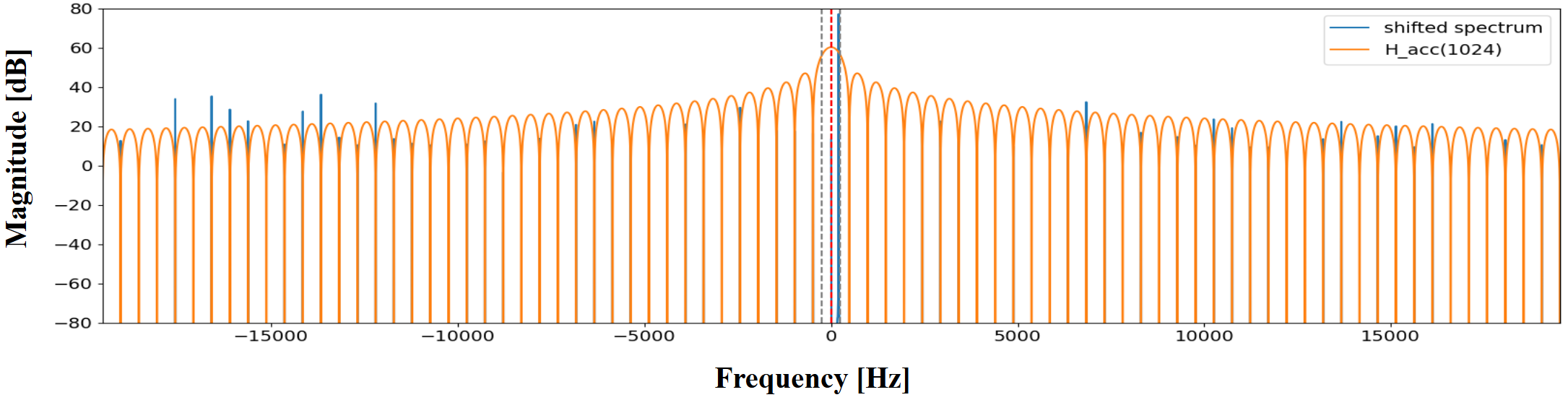}
        \caption{Spectrum after the DDC frequency shift, with the $N_\text{avg}=1024$ averaging-filter (boxcar) response overlaid: the folded comb tones sit in the filter nulls while the band of interest occupies the passband.}
        \label{fig:no_pfb_sim-c}
    \end{subfigure}

    \begin{subfigure}{\textwidth}
        \centering
        \includegraphics[width=0.9\linewidth]{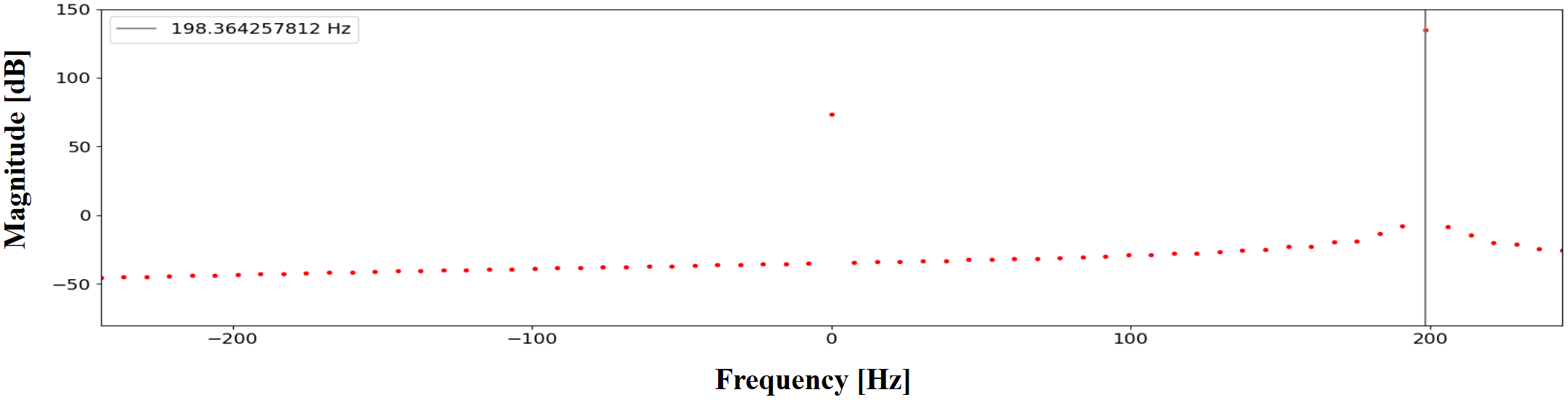}
        \caption{Final decimated output band, 488.28125\,Hz wide and centered at DC, showing only the tone of interest. Here the tone is intentionally placed at 198.364\,Hz (off the $\Delta f_\text{null}$ grid, via a deliberate DDC offset) so the output band is visible away from DC; the residual at DC is 62.3\,dB below the carrier and the broadband floor lies far below it.}
        \label{fig:no_pfb_sim-d}
    \end{subfigure}
    \vspace{1pt}
    \caption{Simulated propagation of a 1024-tone probe comb through the PFB-free receive chain, illustrating how inter-bin aliases are placed in the averaging-filter nulls. Channel~2 is an arbitrary choice shown throughout.}
    \label{fig:no_pfb_sim}
\end{figure}

DSP simulation confirms that this effectively suppresses inter-bin aliases. \Cref{fig:no_pfb_sim} traces a 1024-tone comb across the full 1.024\,GHz band through the receiver DSP chain: the folded tones (aliases) land on the averaging-filter nulls, leaving a decimated output that contains only the tone of interest, its broadband floor far below the carrier (the tone under test is deliberately placed off-grid, at 198.364\,Hz, so that the output carrier is visibly away from DC). The measured RF loopback spectrum of \Cref{fig:output_spectrum_loopback} reproduces the same clean output band.

\begin{figure}[htbp]
  \centering
  \vspace{1pt}
  \includegraphics[width=\textwidth]{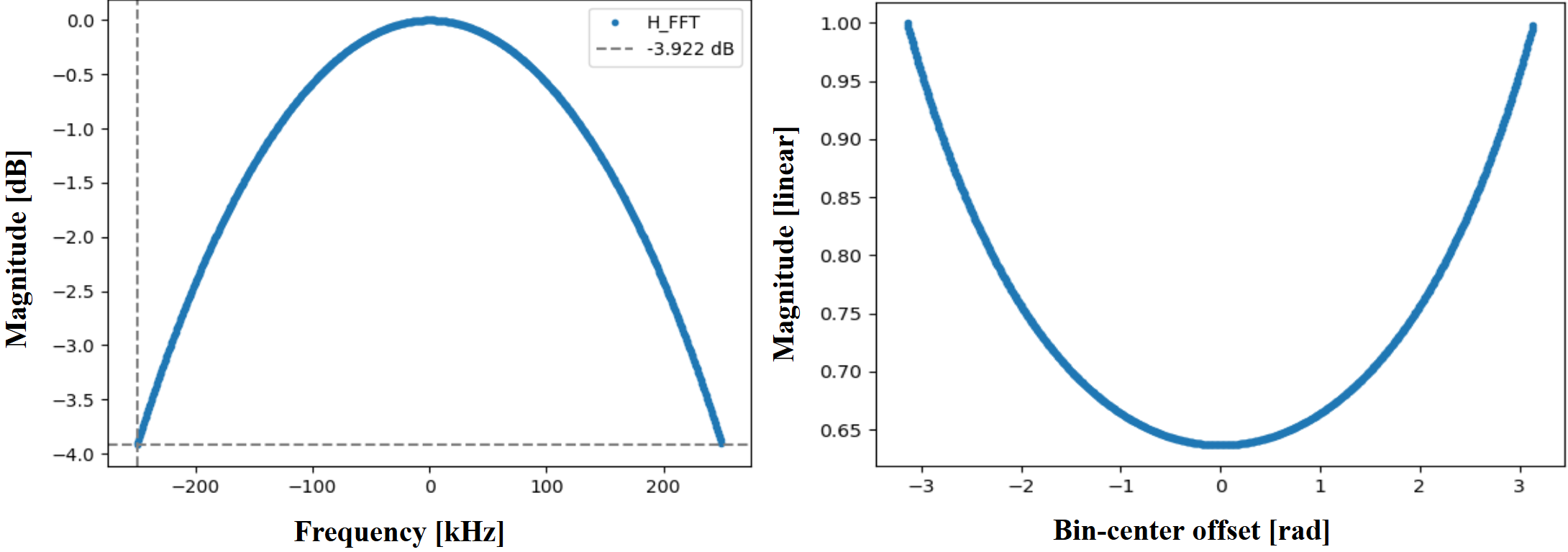}
  \vspace{1pt}
  \caption{FFT-bin passband and its software correction.
    \emph{Left:} Simulated FFT-bin magnitude response $H$ (dB) versus
    frequency offset from bin center; it follows the boxcar
    $\operatorname{sinc}$ envelope, flat at $0$\,dB at bin center and rolling off to
    the annotated $-3.922$\,dB at the $\pm250$\,kHz bin edges ($\pm f_\text{bin}/2$).
    \emph{Right:} The resulting per-tone amplitude correction, plotted against the
    tone's offset from bin center (expressed as the per-bin beat phase: zero at center,
    $\pm\pi$ at the bin edges). The correction normalizes every tone down to the
    worst-case bin-edge level (unity at the edges and $2/\pi \approx
    0.637$ at bin center), pulling a bin-center tone down by $3.922$\,dB onto the
    common level.}
  \label{fig:fft_passband}
\end{figure}

\textbf{Passband flatness.}
A tone placed off-center in its bin is attenuated by the bin's sinc roll-off. The FFT-bin frequency response follows $\operatorname{sinc}(f/f_\text{bin})$, so at worst a tone at bin edge ($f = f_\text{bin}/2$) is attenuated by
\begin{equation}
  20\log_{10}\!\left| \operatorname{sinc}\!\left(\tfrac{1}{2}\right) \right|
    = 20\log_{10}\!\left( \frac{2}{\pi} \right)
    = -3.92\,\text{dB}.
  \label{eq:binedge}
\end{equation}
Because the $\Delta f_\text{null}$ grid does not align with FFT bin centers, probe tones experience up to about 4 dB of inter-channel variations in attenuation if left uncorrected. We calibrate this effect with per-tone scalar corrections computed once at tone-assignment time. Since every tone's bin-center offset is one of $N_\text{avg}=1024$ grid values, the corrections form a 1024-entry LUT applied to the I/Q outputs in software, adding no FPGA resources. The correction normalizes every tone down to the worst-case bin-edge level rather than up to bin center level: scaling down avoids inflating a tone's apparent power, and applying the same scalar to signal and noise leaves the per-channel SNR unchanged. \Cref{fig:fft_passband} shows the simulated FFT-bin response and the resulting correction LUT.

This correction is important for VNA sweeps such as in \Cref{fig:vna_sweep}, where an equal-spaced unit amplitude tone comb is swept across a bandwidth to trace a network's forward transmission coefficient $S_{21}$. Without correction, the FFT-bin $\operatorname{sinc}$ response modulates the measured $S_{21}$, distorting the transmission profile. Removing the plain FFT bin's frequency response therefore yields a more faithful representation of $S_{21}$.

The alias-nulling argument assumes aliases fall exactly within the narrow boxcar nulls. In practice, clock jitter, finite frequency resolution, and imperfect operating conditions can push an alias slightly off its null, so suppression is strong but not perfect; the residual is visible in the crosstalk measurement of \Cref{sec:results:crosstalk}, where an empty neighboring channel's output power stays a few dB above the zero-tone reference. The RF loopback measurements (\Cref{fig:output_spectrum_loopback,fig:phase_noise_loopback}) nonetheless show no aliases or elevated noise floor relative to the PFB-equipped Gen1 baseline.

\section{FPGA IMPLEMENTATION}
\label{sec:implementation}

The Gen2 DSP gateware processes four samples per FPGA clock, whereas Gen1 processes two, both at a 256\,MHz clock rate. Compared with Gen1, this parallel-by-four datapath doubles the sample throughput, and hence the instantaneous bandwidth, without raising the clock frequency. Memory buffer structures throughout the design were widened accordingly, and timing was re-verified in Vivado. Doubling the datapath width does not, however, translate into a uniform doubling of resource usage: the Gen2 transmitter is a novel, DSP- and memory-intensive architecture, and is the dominant driver of the resource and power differences reported below.

\Cref{tab:resources} reports Vivado post-implementation resource utilization for the four-chain, 2048-tone-capable build (gateware tag \texttt{gw-v16p0})\footnote{Gen2 gateware: \url{https://github.com/mx956/primecam_gateware_design2}} on the ZCU111 RFSoC. Gen1 utilization is reported by the current Gen1 Vivado project maintained by the first author.

\begin{table}[H]
\centering
\caption{Vivado post-implementation FPGA resource utilization on the Xilinx ZCU111 RFSoC.
         Gen1 values are from the current Gen1 gateware project maintained by the first author.}
\label{tab:resources}
\vspace{6pt}
\begin{tabular}{lrrrr}
\hline
Resource & Gen1 & Gen2 & Available & Gen2 util.\ (\%) \\
\hline
LUT  & 161{,}167 & 299{,}166 & 425{,}280 & 70.4 \\
FF   & 217{,}352 & 421{,}282 & 850{,}560 & 49.5 \\
BRAM & 169.5     & 1{,}008   & 1{,}080   & 93.3 \\
URAM & 16        & 47        & 80        & 58.8 \\
DSP  & 211       & 1{,}544   & 4{,}272   & 36.1 \\
\hline
\end{tabular}
\end{table}

Look-up tables (LUT) and flip-flops (FF) scale by roughly $1.9\times$ from Gen1 to Gen2, consistent with the wider datapath. Block RAM (BRAM) and DSP slices, by contrast, scale by $5.9\times$ and $7.3\times$ respectively. This disproportionate growth reflects the Gen2 transmitter: where Gen1 streams a pre-computed waveform from a DDR4 look-up table, Gen2 actively synthesizes the probe comb in fabric via the OC-PSB, making the transmitter DSP- and memory-intensive. This active-synthesis transmitter is the principal architectural difference between the two generations.

BRAM is the binding resource at 93.3\,\% utilization. Its dominant consumers are two datapath buffers: the Tone-Select buffer, which is unique to the Gen2 transmitter and has no Gen1 counterpart, and the receive bin-select buffer; both, together with the other pipeline buffers, are doubled in depth by the parallel-by-four processing. Both BRAM and URAM are dedicated memory resources in the FPGA fabric, and BRAM is generally preferred by the designers and synthesis tool because it supports nonzero initialization and is more resource-efficient for smaller memories. We expect that a fraction of this BRAM can be remapped to the device's URAM, which retains headroom at 58.8\,\% utilization, to relieve the BRAM bottleneck and improve the scaling margin.

\Cref{tab:power} reports Gen2 board power measured at the AC wall socket, with a power meter placed in series between the wall outlet and the board's power adapter, at successive operational stages up to the full four-chain, $4\times1024$-tone load.

\begin{table}[H]
\centering
\caption{Measured Gen2 board power at the AC wall socket, by operational stage,
         up to the full four-chain $4\times1024$-tone load. The full-load 82\,W
         corresponds to 20\,mW per KID (4096 tones), or 40\,mW per
         two-KID pixel.}
\label{tab:power}
\vspace{6pt}
\begin{tabular}{lr}
\hline
Operational stage & Power (W) \\
\hline
Powered, idle (no gateware)  & ${\sim}20$ \\
gateware loaded              & ${\sim}45$ \\
Readout started              & ${\sim}56$ \\
\quad + chain 1 (1024 tones) & ${\sim}62$ \\
\quad + chain 2 (1024 tones) & ${\sim}68$ \\
\quad + chain 3 (1024 tones) & ${\sim}75$ \\
\quad + chain 4 (1024 tones) & ${\sim}82$ \\
\hline
\end{tabular}
\end{table}

Idle (powered without gateware upload) board power is about 20\,W. Loading the gateware adds 25\,W, starting the readout DSP chains add a further 11\,W, and each additional 1024 active tones per chain adds 6--7\,W, reaching 82\,W with all four chains loaded. At this full load the 82\,W is spread across 4096 active tones: about 20\,mW per KID, or 40\,mW per dual-polarization pixel. For reference, the Gen1 readout draws under 40\,W at full four-channel operation.\cite{sinclair2023} Gen2's roughly two-fold increase in power is consistent with expectation: relative to Gen1, it doubles the instantaneous bandwidth and datapath width and, through active comb synthesis, substantially increases the processing workload.

At the maximum load ($4\times1024$ tones), we have observed intermittent power-related fault events: the board enters a protection or error state in which SSH access is lost, and normal operation is restored only by power-cycling the board. During these events, the PS and PL temperatures reported by the RFSoC is around $70^{\circ}$C, well below the failure threshold of either, so we suspect a localized temperature or current-draw issue rather than a global over-temperature or over-power condition. The exact root cause is under investigation, and reducing the overall power draw and resource utilization is an active development priority.

\section{MEASURED PERFORMANCE}
\label{sec:results}

\subsection{RF loopback noise measurements between Gen1 and Gen2}
\label{sec:results:noise}

Noise performance is measured in RF loopback configuration where DACs are directly connected to ADCs via coaxial cables. Both Gen1 and Gen2 produced an identical comb of 1000 probe tones at 510\,kHz spacing, with each tone frequency snapped to the nearest 488.28125\,Hz frequency grid. 60 seconds of I/Q time-stream data are captured from Ethernet packets. The test comb spans approximately 510\,MHz because Gen1 is limited to 512\,MHz of instantaneous bandwidth. Using an identical comb for both generations ensures apple-to-apple comparisons between the two.

\begin{figure}[htbp]
  \centering
  \includegraphics[width=\textwidth]{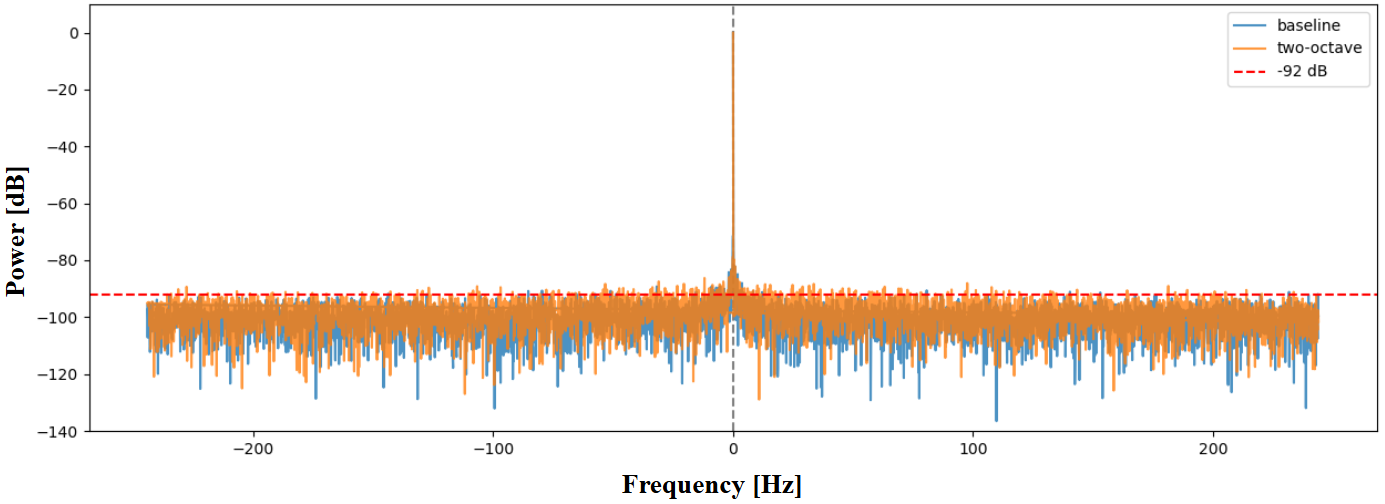}
  \vspace{1pt}
  \caption{Normalized RF loopback output power spectrum from an arbitrary probe tone channel.
    Both Gen1 (baseline) and Gen2 (two-octave) show a single
    carrier peak at 0\,Hz and a flat noise floor at $-92$\,dBc
    with no spurious aliased components across the $\pm244.140625$\,Hz output band.
    The absence of aliases confirms that per-tone 488.28125\,Hz grid
    alignment places all inter-bin aliases in the averaging nulls.}
  \label{fig:output_spectrum_loopback}
\end{figure}

\Cref{fig:output_spectrum_loopback} shows the normalized output power spectrum of a probe tone channel of typical performance. Both Gen1 and Gen2 exhibit a single carrier peak at 0\,Hz with a flat noise floor at $-92$\,dBc across the full $\pm244.140625$\,Hz output band without visible spurious peaks. This validates the alias-suppression scheme: a failure of the 488.28125\,Hz grid constraint would produce spurious aliased components in the output band, or an elevated broadband noise floor if many aliases fold in at irregular frequency offsets. The clean spectrum shows neither.

\begin{figure}[htbp]
  \centering
  \includegraphics[width=\textwidth]{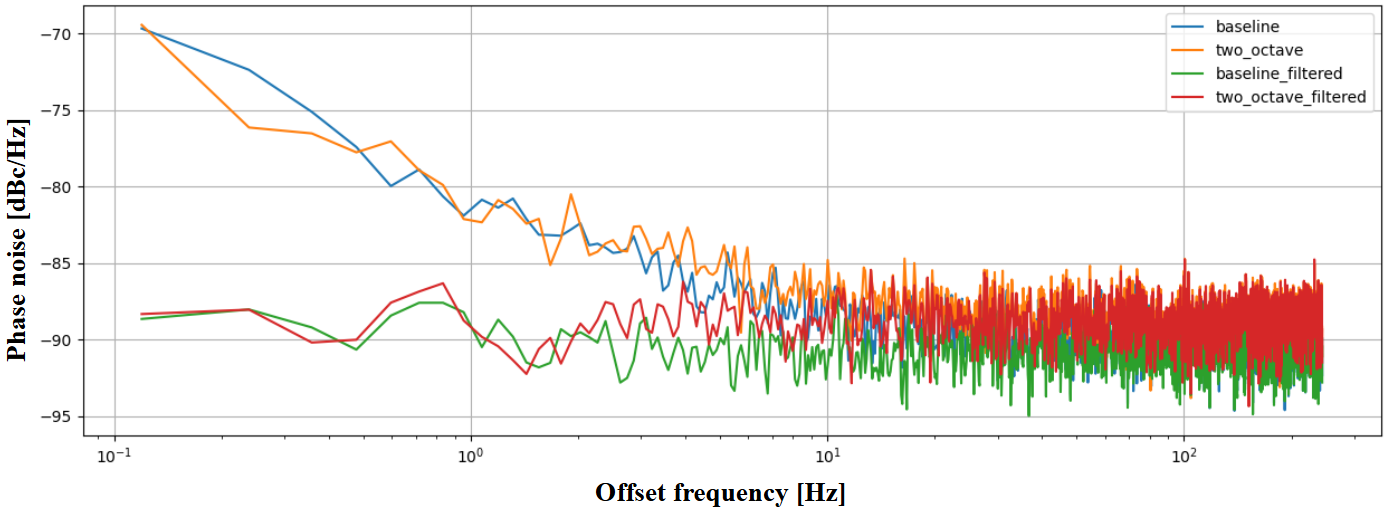}
  \vspace{2pt}
  \caption{Single-sideband phase noise spectra from an RF loopback test
    (DAC output to ADC via coaxial cable). Plotted from the same data as in \Cref{fig:output_spectrum_loopback}.
    Baseline: Gen1 receive chain with polyphase analysis filter bank.\cite{sinclair2022}
    Two-octave: Gen2 receive chain with plain 2048-point FFT (this work).
    Filtered variants subtract the first five PCA modes computed across
    the 1000-tone ensemble, greatly suppressing common-mode $1/f$ noise.
    The Gen1 and Gen2 traces overlap and are indistinguishable across the full
    range in both the unfiltered and PCA-filtered cases, indicating comparable noise performance between Gen1 and Gen2.}
  \label{fig:phase_noise_loopback}
\end{figure}

\Cref{fig:phase_noise_loopback} shows the corresponding single-sideband phase-noise spectra. The Gen1 and Gen2 traces overlap and are essentially indistinguishable across the full 0.1--244\,Hz range in both the unfiltered and filtered cases. The filtered results are produced by subtracting the first five principal components (PCA) across the 1000-tone ensemble to suppress common-mode $1/f$ noise.

\subsection{On/off-resonance noise measurements with Gen2 readout}
\label{sec:results:detector}

Following the dark-measurement method of Sinclair et al.~\cite{sinclair2024noise}, we assess whether the Gen2 readout is detector-noise-limited by comparing each channel's noise on and off resonance. On resonance, the noise is the sum of detector and readout contributions; off resonance, with the tones moved $1$\,MHz away (many linewidths, which are of order $10$\,kHz for FYST-class arrays~\cite{sinclair2024noise,vaskuri2025}), the detector contribution is negligible and only the readout noise remains~\cite{sinclair2022}. We assume that the readout noise is unchanged with the 1\,MHz frequency shift. By ``readout noise'' we mean the noise from the entire readout chain except the detector's contribution: the DSP; the RFDC data converters; the warm attenuators and amplifiers; the cryogenic low-noise amplifier; the off-resonance detector chip and feedline; and the cables. Because uncorrelated noise powers add, an on/off noise power ratio above a factor of two ($3$\,dB) means the detector noise dominates the readout noise (detector-noise-limited), while a ratio below $3$\,dB means readout-noise-limited operation~\cite{sinclair2024noise}.

The cryogenic RF setup and the readout procedure broadly follow the Gen1 deployment of Patel et al.~\cite{patel2025ccatreadout10000280} with a less mature configuration. The warm-side RFSoC readout differs in the Gen2 system of \Cref{sec:dsp,sec:implementation}. Because Gen2 is planned to be software-compatible with the Gen1 infrastructure, the same procedures apply and we omit a detailed description. Unlike Patel et al.~\cite{patel2025ccatreadout10000280}, which derives frequency and dissipation quadrature noise from a resonator circle fit; and Sinclair et al.~\cite{sinclair2024noise}, which sums the individual I and Q noise power spectral densities (PSD), we compute phase noise directly from the raw I/Q timestream: the per-tone phase $\phi = \arg(I + jQ)$ is computed and its single-sideband PSD is estimated by Welch's method in dBc/Hz. With small phase fluctuations (which is true when dark), the phase noise of a probe tone placed on resonance should contain mostly the frequency noise (noise in the direction of the I/Q circle), which is the relevant detector noise, as opposed to the dissipation noise (noise normal to the I/Q circle).

We perform this measurement using two test chips: a populated two-octave chip and a two-level-system noise (TLS-noise) chip. The two-octave chip, connected to readout chain 3 (ch3), contains KIDs similar to those in the science-grade 850\,GHz arrays. The TLS-noise chip, connected to ch4, contains two sets of KIDs with different architectures originally intended for TLS-noise studies. Four independent readout chains, each connected to a test chip, are installed in the cryostat to provide a large KID sample size. However, only ch3 and ch4 showed valid transmission at the time of measurement, so the results reported here are limited to those two chains.

\begin{figure}[htbp]
  \centering
  \includegraphics[width=0.499\textwidth]{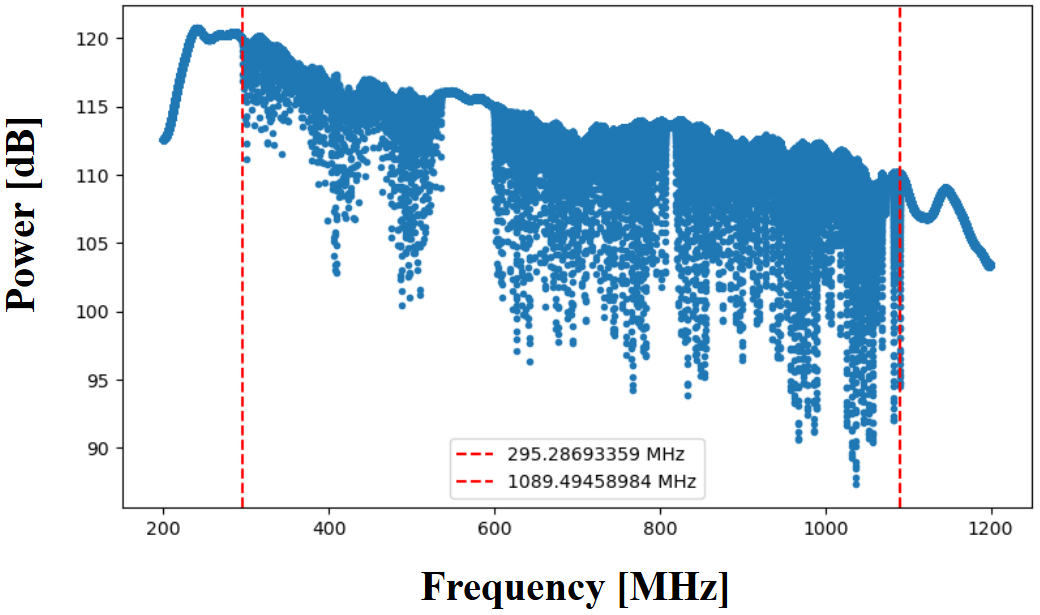}\hfill
  \includegraphics[width=0.499\textwidth]{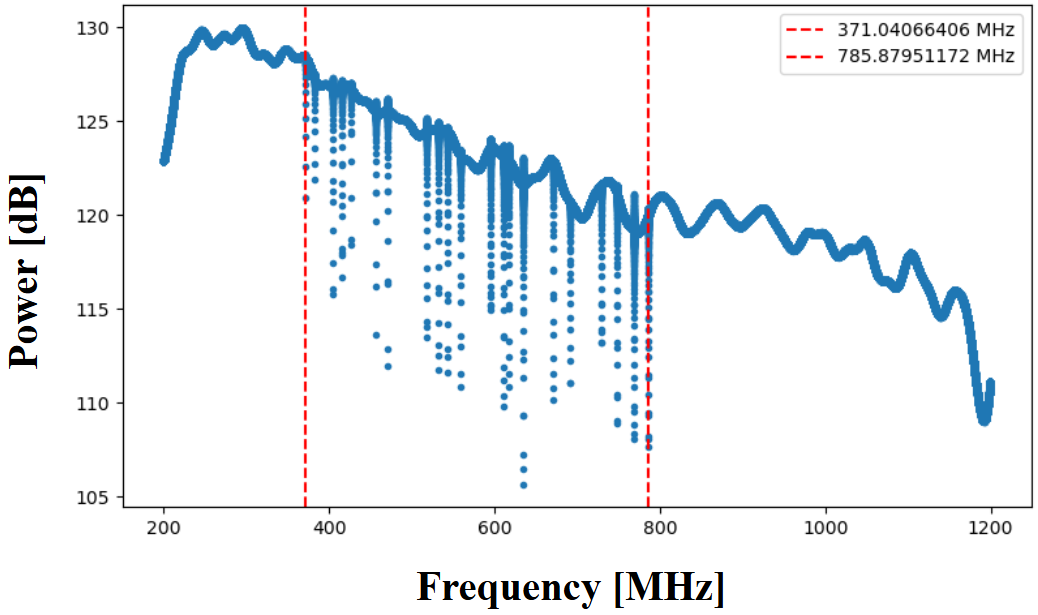}
  \vspace{2pt}
  \caption{Full-band VNA sweeps (forward transmission $S_{21}$ versus\ frequency, in
    dB) across the 1.024\,GHz readout band centered at 700\,MHz, each acquired in a
    single sweep. \emph{Left:} ch3, the populated two-octave chip, with
    272 software-identified resonator dips; \emph{Right:} ch4, the half-cap
    TLS-noise chip, with 22 software-identified resonator dips. Vertical dashed lines mark the
    resonant frequencies of the lowest and the highest frequency resonators identified
    by the readout system.}
  \label{fig:vna_sweep}
\end{figure}

Each chain was first measured with a full-band VNA sweep. \Cref{fig:vna_sweep} shows the forward transmission ($S_{21}$) across the full 1.024\,GHz band centered at 700\,MHz; the resonator-finding software identifies 272 dips on ch3 and 22 on ch4. The whole band is acquired in a single sweep, exercising the two-octave capability of Gen2.

\begin{figure}[h]
  \centering
  \includegraphics[width=0.499\textwidth]{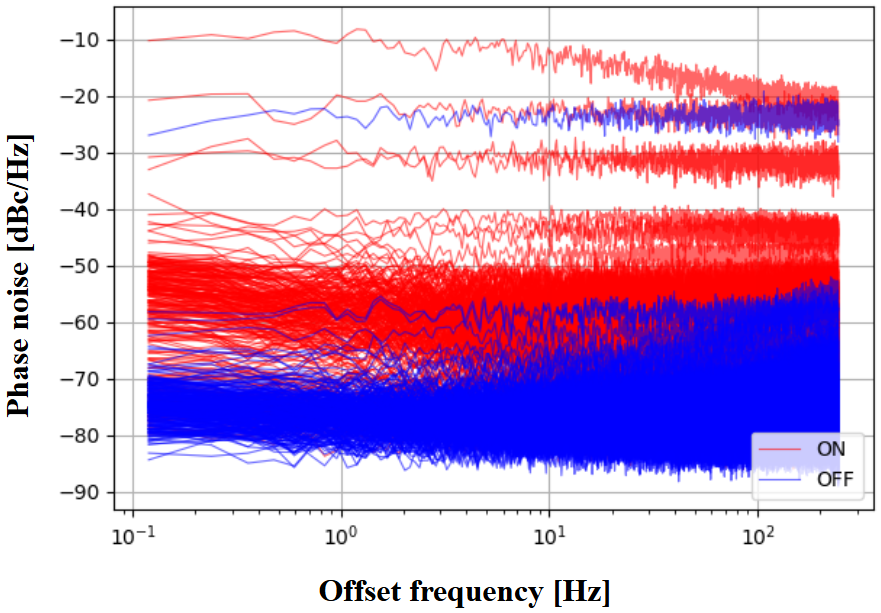}\hfill
  \includegraphics[width=0.499\textwidth]{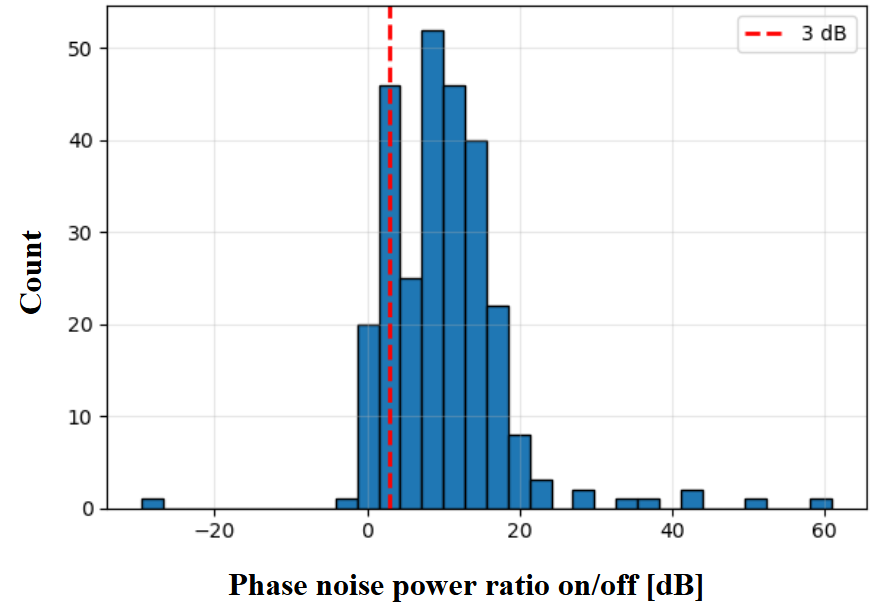}\\[2pt]
  \includegraphics[width=0.499\textwidth]{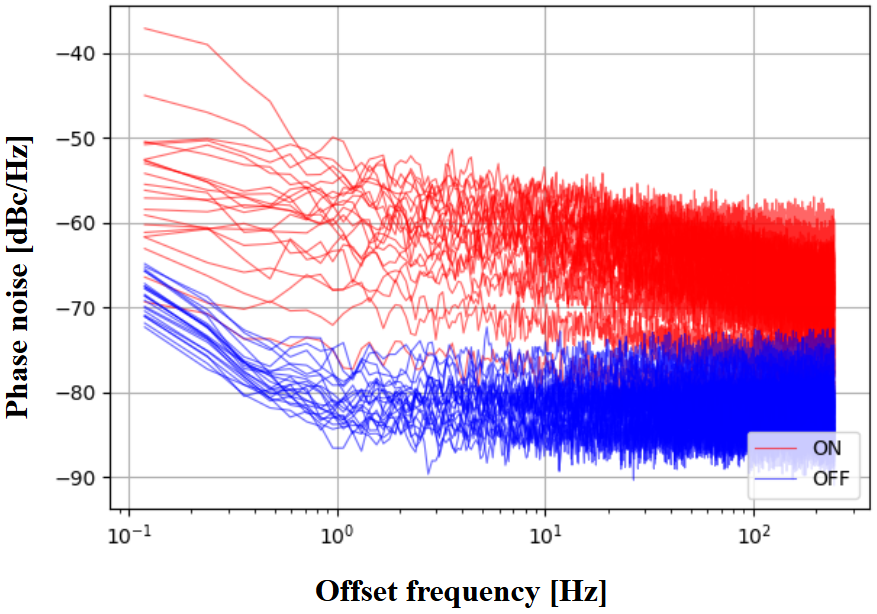}\hfill
  \includegraphics[width=0.499\textwidth]{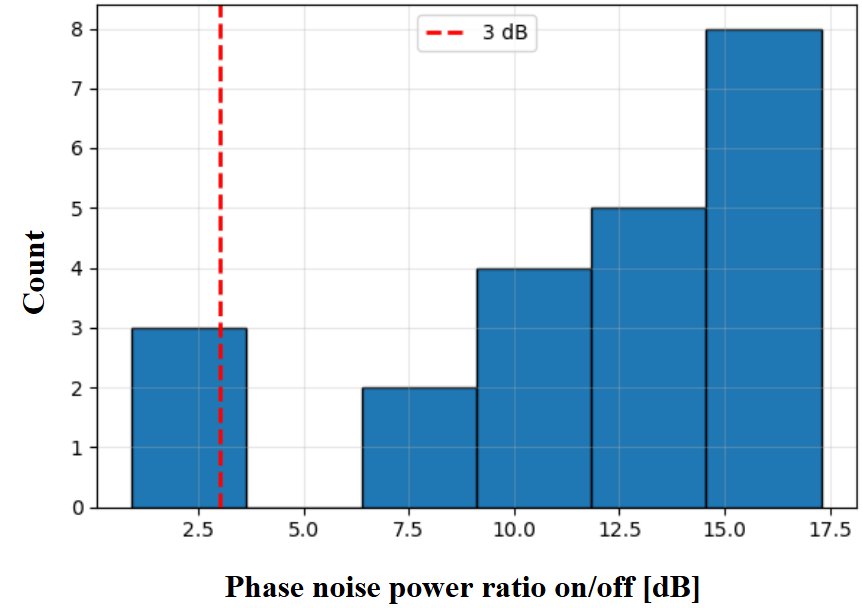}
  \vspace{2pt}
  \caption{On/off-resonance phase noise for two test chips measured with Gen2 readout. \emph{Top:} Populated
    two-octave chip, 272 KIDs. \emph{Bottom:} TLS-noise chip, 22 KIDs. \emph{Left:} Per-channel single-sideband phase-noise spectra, red on-resonance and blue off-resonance. \emph{Right:} Histogram of the per-channel on/off noise-power ratio (median of the high-offset band); the dashed line marks the 3-dB
    detector-noise-limited threshold~\cite{sinclair2024noise}. For both chips, the majority of channels show on/off noise power ratio above 3 dB.}
  \label{fig:detector_noise}
\end{figure}

\Cref{fig:detector_noise} presents the on/off-resonance measurement of two test chips by Gen2 readout with per-channel on- and off-resonance phase-noise spectra (left) and a histogram of the per-channel on/off power ratio, taken as the median over the high-offset white-noise band, with the 3-dB threshold annotated (right). For both test chips, the large majority of channels show on-resonance noise power above the off-resonance noise power with the ratio above 3 dB, indicating detector-noise-limited operation.

These results are preliminary: the detectors were measured dark, without the optical loading they are designed for; the per-channel tone powers were not optimized; and the timestreams are uncalibrated raw I/Q. Even so, the on/off ratios suggest detector-noise-limited performance for a large fraction of channels, matching the Gen1 findings of Sinclair et al.\cite{sinclair2024noise}. More comprehensive measurements using a cryogenic blackbody, optimized per-resonator biasing, and calibrated I/Q timestreams will be performed as the Gen2 software integration and the 850\,GHz test cryostat setup continue to develop.

\subsection{Digital channel crosstalk}
\label{sec:results:crosstalk}
Crosstalk between channels adds unwanted signal power and should be measured and bounded. A particular point of concern in our DSP pipeline is where tones share the same nonzero bin-center offset and thus the same DDC frequency shift. We predict that crosstalk in this condition can be reduced by constraining the tones to have distinct bin-center offsets at least one frequency grid step apart.

\begin{figure}[p]
  \centering
  \includegraphics[width=0.8\textwidth]{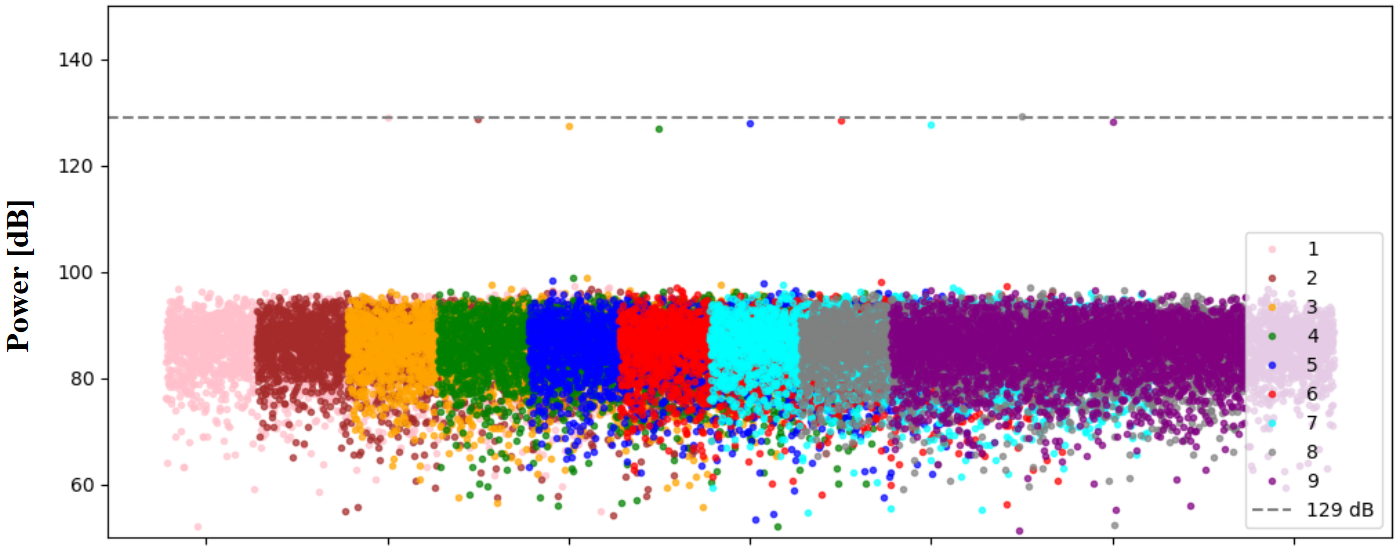}\\[4pt]
  \includegraphics[width=0.8\textwidth]{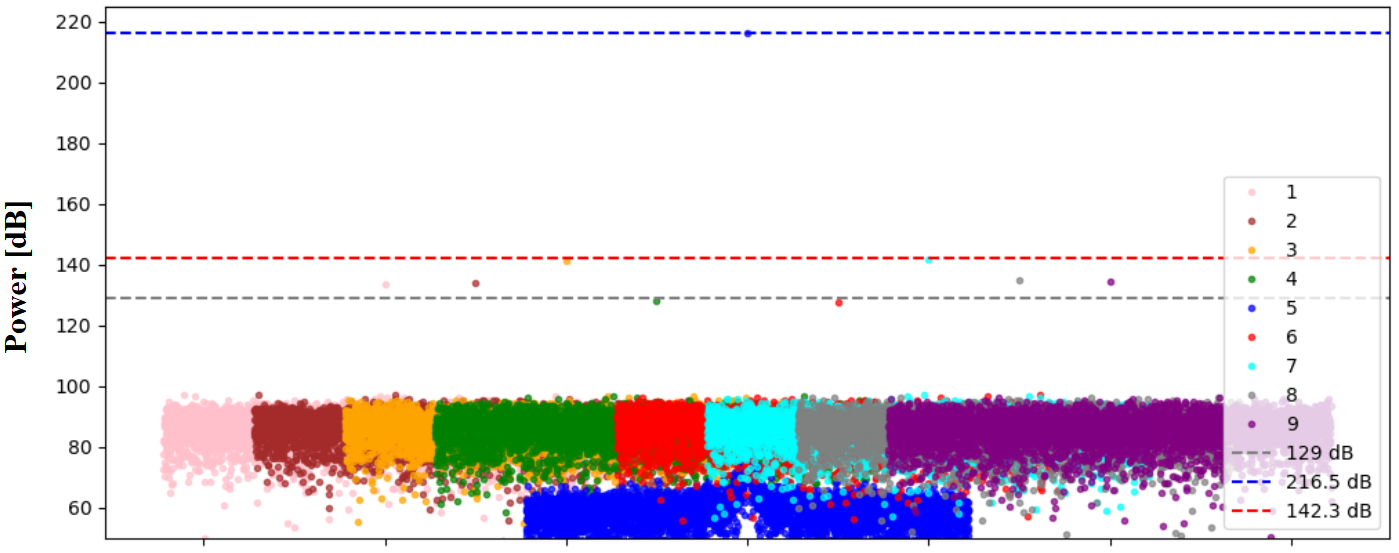}\\[4pt]
  \includegraphics[width=0.8\textwidth]{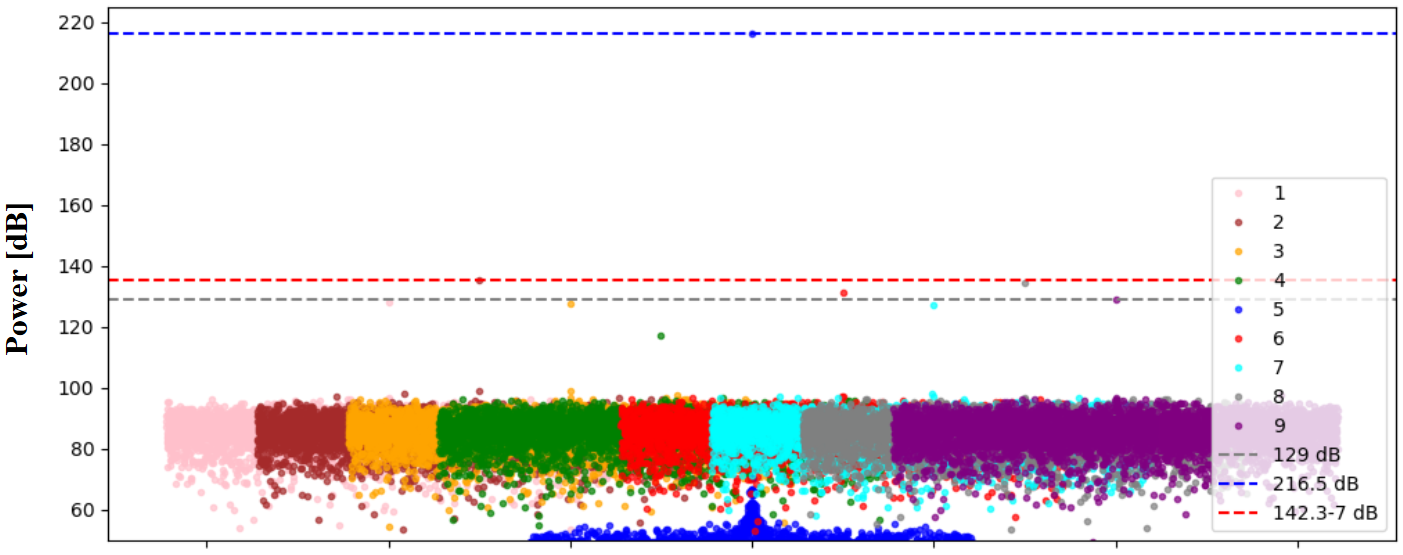}
  \vspace{1pt}
  \caption{Digital channel crosstalk, measured
    with nine tones in nine adjacent FFT bins (probe tone channels). Each panel shows the nine
    per-channel output spectra, displaced side by side by different colors for visibility; the vertical axis is the uncalibrated output power in dB, so levels are meaningful only as differences and are
    quoted in the text as dBc relative to the Channel~5 carrier.
    \emph{Top:} Zero tone power. The residual DC peak sets the reference ($-87.5$\,dBc).
    \emph{Middle:} Channel~5 carrier at full power, all bin-center offsets identical.
    The nearest neighbors' DC power rises $13.3$\,dB above the reference (to $-74.2$\,dBc).
    \emph{Bottom:} Same measurement with neighbor offsets shifted by one output
    bandwidth. The worst-case rise falls to $6.3$\,dB (to $-81.2$\,dBc).}
  \label{fig:bin_crosstalk}
\end{figure}

\Cref{fig:bin_crosstalk} shows the measurement process. The spectra shown are PCA-filtered to suppress common-mode noise. Crosstalk is read from each channel's power at DC, where a neighbor's crosstalk overlays, quoted in dBc relative to the Channel~5 carrier. Starting with zero tone power for all tones (\Cref{fig:bin_crosstalk}a), the residual DC peak sets a common reference at $-87.5$\,dBc. Energizing the tone in Channel~5 only, with all channels' frequency offsets identical, raises the neighbors' DC-bin power (\Cref{fig:bin_crosstalk}b). The largest increase occurs in the two nearest channels, where the power reaches $-74.2$\,dBc, $13.3$\,dB above the zero power reference. The increase tapers with distance. Separating the bin-center offsets by one frequency grid step (\Cref{fig:bin_crosstalk}c) reduces the worst-case rise to $6.3$\,dB above the reference. Separations of two, three, five, and ten grid steps are indistinguishable from one, therefore a single grid step separation in bin-center frequency offsets between tones captures the available crosstalk reduction.

We expect the probability of detector resonances having identical bin-center offsets to be low but non-negligible. As the science-grade 850\,GHz KID arrays are currently under fabrication and characterization, we report this crosstalk without assessing whether it limits science performance at this time. If it proves consequential, a low-cost, software-only mitigation is to enforce a minimum separation of one frequency grid step (488.28\,Hz) between the tones' frequency offsets from their respective bin centers. Because 488.28\,Hz is only about 5\% of a resonator linewidth, which is of order 10\,kHz for large-format CCAT arrays\cite{sinclair2024noise,vaskuri2025}, such an adjustment moves a tone only slightly off resonance.

\section{CONCLUSIONS}
\label{sec:conclusion}
We have presented the Gen2 KID readout system implemented on a Xilinx ZCU111 RFSoC. Designed to meet the readout bandwidth and tone-capacity requirements of the CCAT Prime-Cam 850\,GHz and 410\,GHz instrument modules, the architecture also provides capabilities that may be useful more broadly across Prime-Cam and in other probe-tone-based FDM readout systems. Gen2 comprises four independent readout chains, each with 1.024\,GHz instantaneous bandwidth and support for up to 2048 probe tones. Relative to Gen1, the OC-PSB transmitter doubles the bandwidth and tone capacity while providing runtime control of individual tone frequency, amplitude and phase. The receiver uses a 2048-point FFT without an analysis PFB, with inter-bin aliases suppressed through tone placement on the null grid of the downstream averaging filter.

RF loopback measurements show noise performance comparable to Gen1 baseline readout under the tested conditions, and preliminary dark measurements with two KID test chips indicate detector-noise-limited operation for the majority of measured channels. The current implementation is limited primarily by high BRAM utilization and power-related fault behavior at maximum load. Ongoing work addresses these limitations together with software integration, calibrated optical measurements, and closed-loop tone tracking.

\acknowledgments

The CCAT project, FYST and Prime-Cam instrument have been supported by generous contributions from the Fred M.\ Young, Jr.\ Charitable Trust, Cornell University, Duke University, the Canada Foundation for Innovation, and the Provinces of Ontario, Alberta, and British Columbia. The construction of the FYST telescope was supported by the Gro{\ss}ger{\"a}te-Programm of the German Science Foundation (Deutsche Forschungsgemeinschaft, DFG) under grant INST\,216/733-1 FUGG, as well as funding from Universit{\"a}t zu K{\"o}ln, Universit{\"a}t Bonn, and the Max Planck Institut f{\"u}r Astrophysik, Garching. The construction of EoR-Spec is supported by NSF grant AST-2009767. The construction of the 350\,GHz instrument module for Prime-Cam was supported by NSF grant AST-2117631. The construction of the 850\,GHz instrument module for Prime-Cam is supported by CFI projects 39656 and 46097, and Canadian provincial matching funds. R. Xie acknowledges support and facilities from NRC Herzberg Astronomy and Astrophysics. The completion and deployment of the Prime-Cam instrument with the initial instrument modules is supported by a generous contribution from Alex Gerko, Founder and CEO of XTX Markets.

Drafting this manuscript, the corresponding author used Anthropic's Claude, a large language model, to assist with LaTeX syntax and language clean-up. All technical content, data, analysis, figures, and references were produced and verified by the author, who reviewed AI-assisted text and takes full responsibility for the content of this paper.

\bibliographystyle{spiebib}
\bibliography{references}

\end{document}